\documentclass[twocolumn,twoside]{revtex4}
\pdfoutput=1
\usepackage{graphicx}
\usepackage{fancyhdr}
\def\be{\begin{equation}}
\def\ee{\end{equation}}
\def\bea{\begin{eqnarray}}
\def\eea{\end{eqnarray}}

\newcommand{\beq}{\begin{equation}}
\newcommand{\eeq}{\end{equation}}
\newcommand{\la}{\langle}

\newcommand{\beqa}{\begin{eqnarray}}
\newcommand{\eeqa}{\end{eqnarray}}

\begin{document}

\title{Theory perspectives on rare Kaon decays and CPV}

%

\author{Giancarlo D'Ambrosio\\}
\affiliation{INFN  Sezione di Napoli, Italy}

\begin{abstract}
I review  rare kaon decays in the LHC era: we discuss interplay with B-anomalies and possible New Physics in direct CP violation in $K\to 2\pi$: very rare kaon decays like  $K \to \pi \nu \bar{\nu}$ are very 
important to this purpose.
We discuss also the decays $K^0 \to  \mu ^+ \mu ^-$ due to the LHCB measurement
\end{abstract}

\maketitle

\thispagestyle{fancy}


\section{Introduction and $K \to \pi \nu \bar{\nu}$}
\begin{figure}
 \centerline{\includegraphics[width=0.7\linewidth]{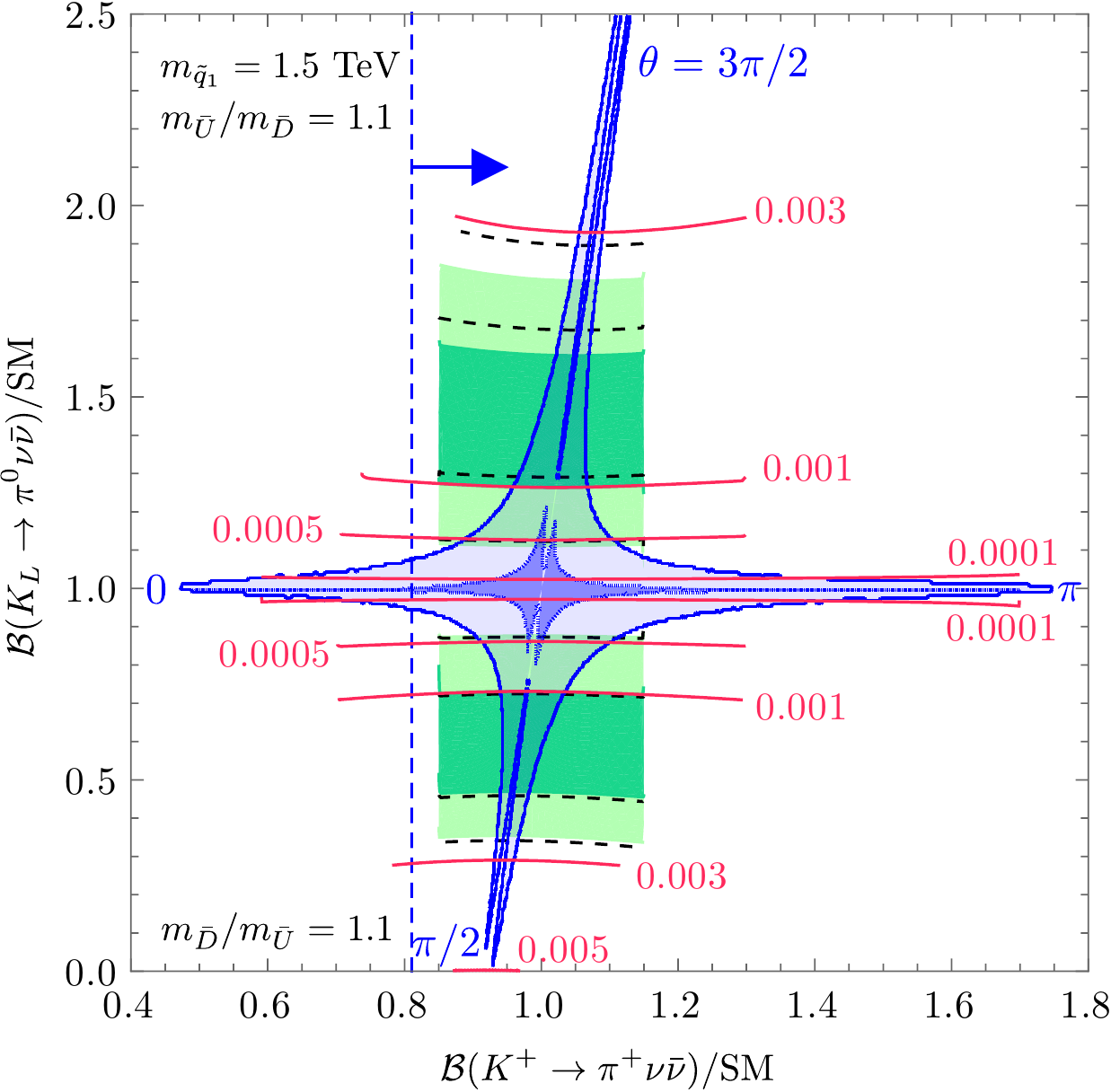}}
 \caption[]{$K \to \pi \nu \bar{\nu}$:  NP from  $K\to 2\pi$ susy  isospin breaking terms (${\Im} (A_2)$)    \cite{Kitahara:2016otd,Crivellin:2017gks}}
 \label{fig:Kpinunu}
\end{figure}

 Rare kaon decays furnish challenging MFV probes and will severely constrain additional flavor physics motivated by NP \cite{Crivellin:2017gks}. 
SM predicts the $V-A\otimes V-A$ effective hamiltonian (Fig. \ref{fig:Kpinunu})
\begin{eqnarray}
{\cal H}& = \frac{G_{F}}{\sqrt{2}} \frac{\alpha}{2 \pi \sin ^2 \theta_W} \overline{
s}_L \gamma _\mu d_L \  \overline{\nu } _L \gamma ^\mu \nu _L  \times
\nonumber\\ &\nonumber\\ &
(\ \underbrace{V_{cs}^{*}V_{cd}\ X_{NL}}_{\textstyle{\lambda x_c}} \,
+ \underbrace{ V_{ts}^{*}V_{td}X(x_t)}_{\textstyle{A^2\lambda ^5 \ 
(1-\rho -i{\eta}){ x_t }}})  \label{ampsd} 
\end{eqnarray}
 $x_q=m_q^2/M_W^2$, $\theta_W$ the Weak angle and 
 $X$'s are the Inami-Lin functions with 
Wilson coefficients known at two-loop electroweak corrections  and 
the main uncertainties
is due to the strong corrections in the charm loop contribution.
The structure in (\ref{ampsd}) leads to a pure CP violating contribution
to $K_{L}\rightarrow \pi ^{0}\nu \overline{\nu },$ induced only from the top
loop contribution and thus proportional to $\Im m(\lambda _{t} )$
($\lambda _t= V_{ts}^{*}V_{td}$) and free of
hadronic uncertainties. This leads to the SM prediction$$ K_L  \quad    =   (2.9\pm 0.2 ) \times 10^{-11}\quad  K^+ \quad   = (8.3\pm 0.9 ) \times 10^{-11}. \label{eq:smk}$$
where   the   parametric uncertainty due to  the error on 
$|V_{cb}|$, $\rho$ and $\eta$ is shown.

\begin{figure}
\begin{minipage}{0.46\linewidth}
\centerline{\includegraphics[width=1.4\linewidth]{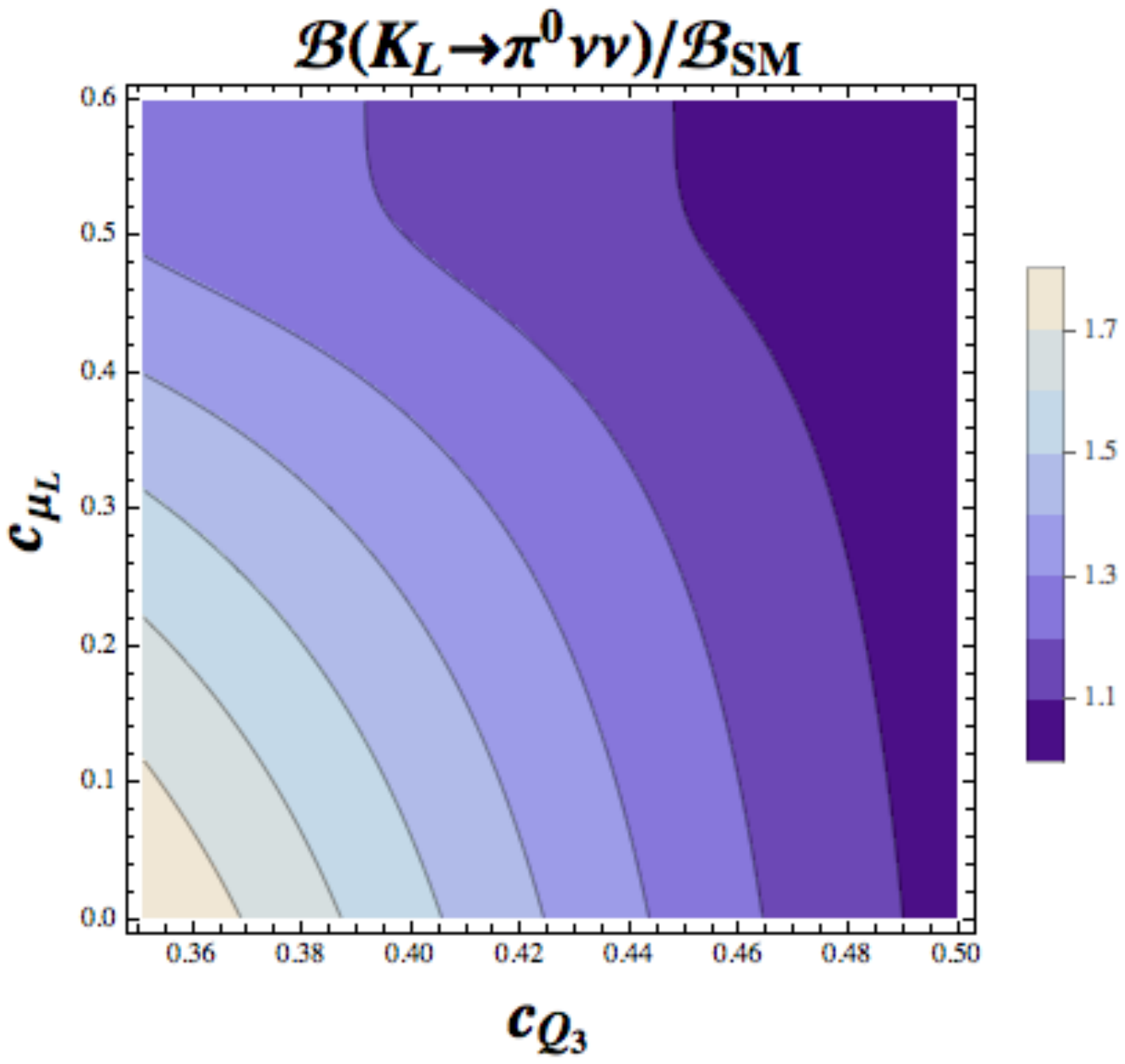}}
\end{minipage}
\begin{minipage}{0.46\linewidth}
\centerline{\includegraphics[width=1.4\linewidth]{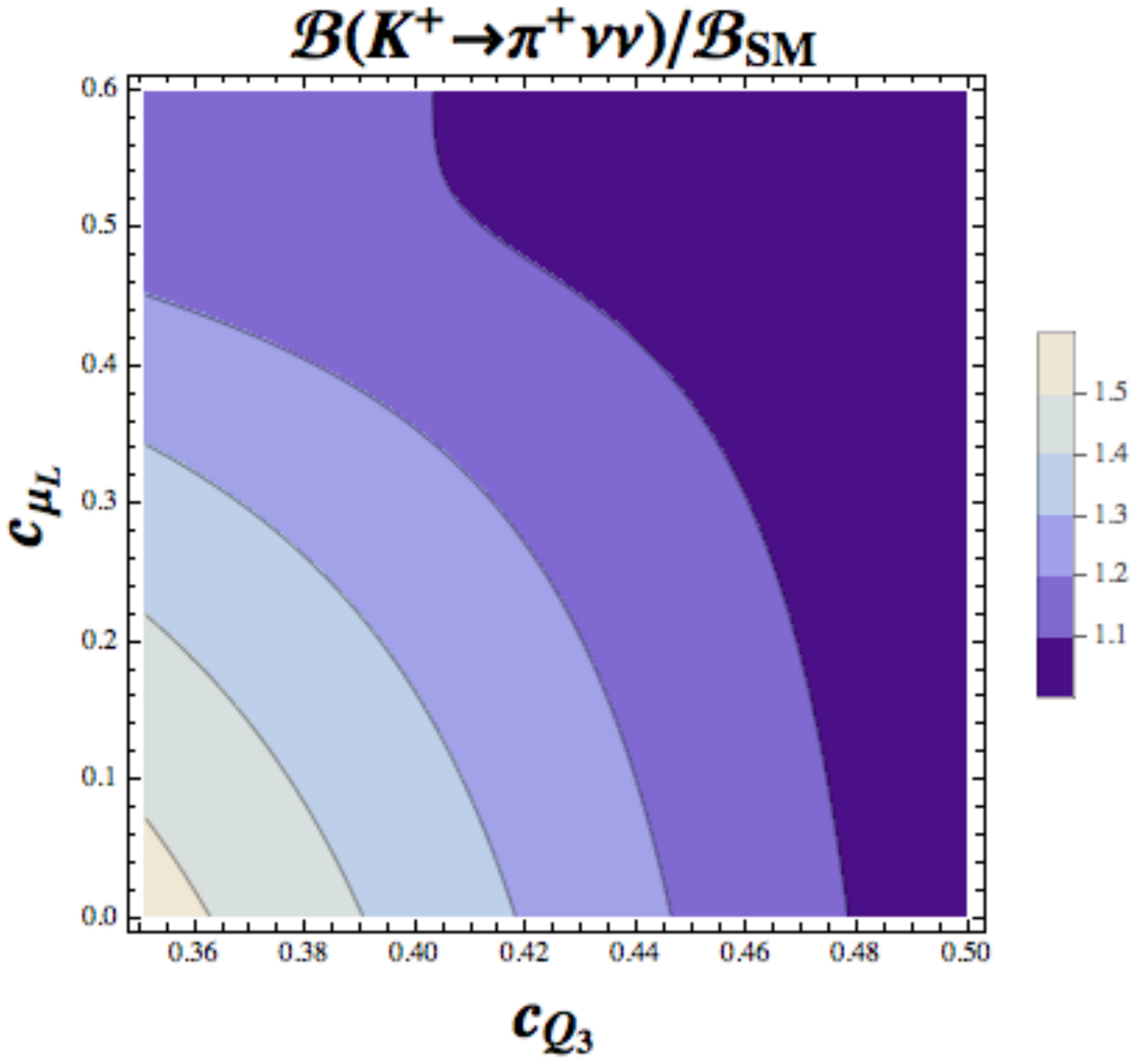}}
\end{minipage}
\caption[]{RS scenario to explain B-anomalies:    $B(K \to \pi \nu \bar{\nu})$ ranges 
as a function of fermion profiles ($c_i$'s)}
\label{fig:extradim}
\end{figure}

Typical BSM predict new flavor structure that might affect  $K \to \pi \nu \bar{\nu}$ that now can be tested  at   NA62 and KOTO {\cite{Mirra}}; we describe two different   BSM  effects
i) new flavor structures  for $\epsilon '$ avoiding    $\Delta S=2$ constraints 
(Fig. \ref{fig:Kpinunu}) \cite{Kitahara:2016otd,Crivellin:2017gks}
 and ii) attempts to  describe B-anomalies \cite{Hurth:2017sqw}, typically induce large flavor effects at O${(1)}$ \ TeV \cite{DAmbrosio:2017wis}.
i) the recent  lattice results for $K\to 2\pi$   leave  open the possibility of BSM for $\epsilon '$; to isospin breaking terms in ${\Im} (A_2)$  have been studied \cite{Kitahara:2016otd} in Fig.\ref{fig:Kpinunu}. We expect effects at most 10\% in  $K ^+ \to \pi ^+ \nu \bar{\nu}$
while are more sizable for $K _L\to \pi ^0 \nu \bar{\nu}$.
 Theoretically addressing  flavor  in Randall Sundrum  models is more challenging: we have studied the so called flavor anarchy scenario with 5D MFV  and custodial symmetry; the only sources of flavor breaking are two 5D anarchic Yukawa matrices. These matrices also generate also the bulk masses, which are responsible for the resulting flavor hierarchy. The theory flows to a next to minimal flavor violation model where flavor violation is dominantly coming from the 3rd generation.  We show that it is possible to find a range of parameters for 
bulk masses satisfying experimental flavor constraints, but also we explain the neutral B-anomalies, requiring NP flavor  scale at  O${(1)}$ \ TeV.
Then we  
 address  $K   \to \pi   \nu \bar{\nu}-$decays: we show the TH predictions as a function of the bulk fermion masses   in  Fig.\ref{fig:extradim}  \cite{DAmbrosio:2017wis}. A natural issue is to test  O${(1)}$ \ TeV physics at LHC; we are trying  to apply the technique of Ref.  \cite{Chakraborty:2017mbz}  to this purpose.

\section{$K_{L,S}\rightarrow \mu ^{+}\mu ^{-}$} 
Recent $K_S \rightarrow\mu\overline{\mu}$  LHCB measurement
is very interesting and unexpected
\begin{eqnarray}
B(K_S \rightarrow\mu\overline{\mu})_{LHCB}< 9\times 10^{-9} \  {\rm at} \  90    \  \% \   {\rm CL} \\
B(K_S \rightarrow\mu\overline{\mu})_{SM}=(5.0\pm 1.5) \times 10^{-12}.
\label{eq:KSmumuexp}\end{eqnarray}
It represents an important milestone since   it has improved the previous limit, 
$<3.2\times 10^{-7}\  {\rm at} \  90    \  \% \   {\rm CL}$,  lasted 40 years. It is based on a production of $10^{13} \ K_S $ per fb$^{-1}$ inside the LHCB acceptance and it  is obtained using 1.0 fb$^{-1}$ of pp collisions at $\sqrt{s} = 7 \ {\rm TeV}$ collected   in 2011.

Two photon exchange generates the dominant contribution for both $K_L$ and $K_S$ decays to two muons \cite{DAmbrosio:2017klp}. The structure of weak and electromagnetic interactions entails a vanishing CP conserving short distance contribution to $K_{S}\rightarrow \mu ^{+}\mu ^{-}$.  Indeed the  SM short diagrams (similar to  $K   \to \pi   \nu \bar{\nu}$ in Fig. 1) lead to  the SM effective hamiltonian similar to eq. (\ref{ampsd}).
 
\begin{figure}
\begin{minipage}{0.45\linewidth}
\centerline{\includegraphics[width=0.8\linewidth]{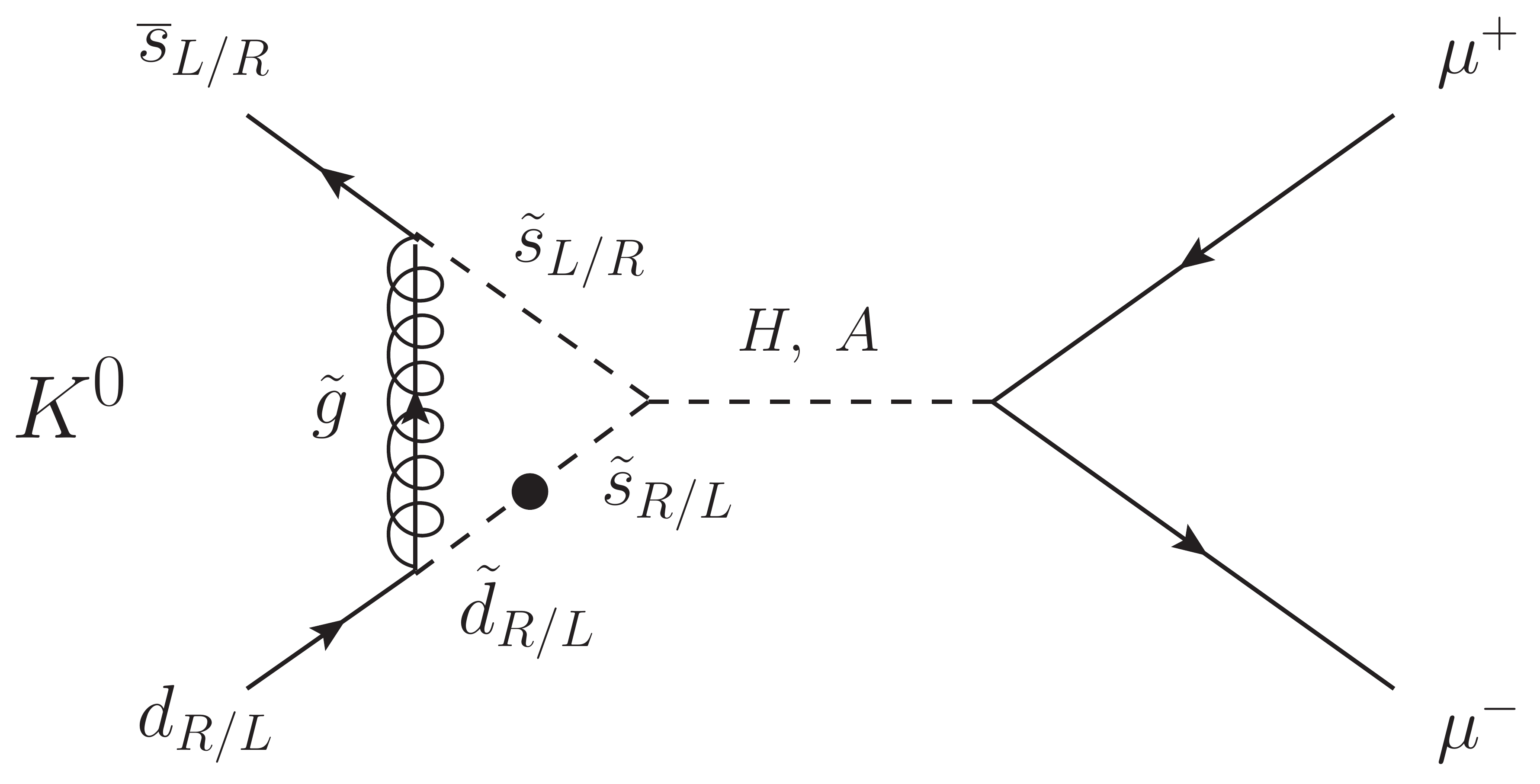}}
\end{minipage}
\hfill
\begin{minipage}{0.5\linewidth}
\centerline{\includegraphics[width=0.8\linewidth]{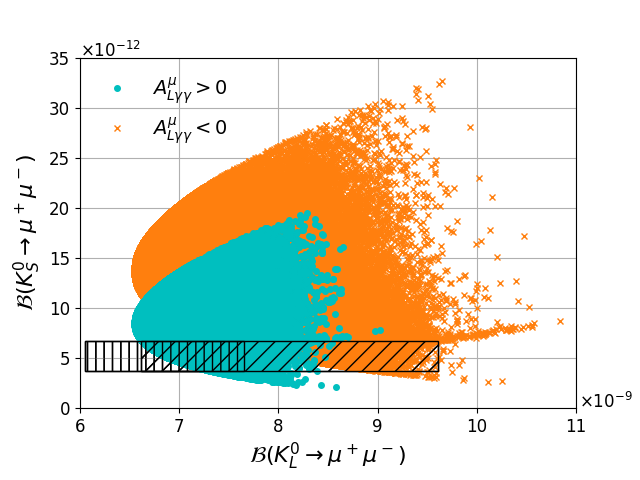}}
\end{minipage}
\caption[]{Susy scenario: $K_S\to \mu \mu$ diagram (left), theory predictions:  in dashed area no  interference effects are considered  (right)}
\label{fig:KSsusy}
\end{figure}

\begin{figure}
\begin{minipage}{0.46\linewidth}
\centerline{\includegraphics[width=1\linewidth]{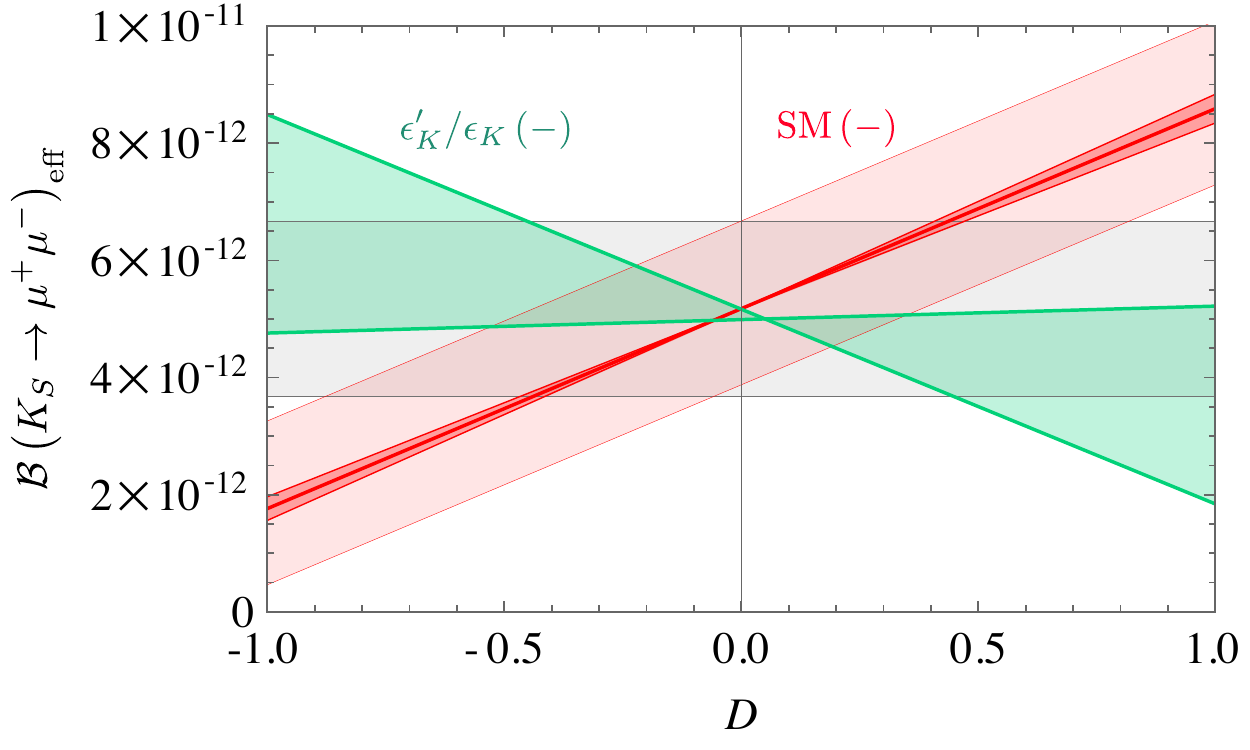}}
\end{minipage}
\hfill
\begin{minipage}{0.47\linewidth}
\centerline{\includegraphics[width=1\linewidth]{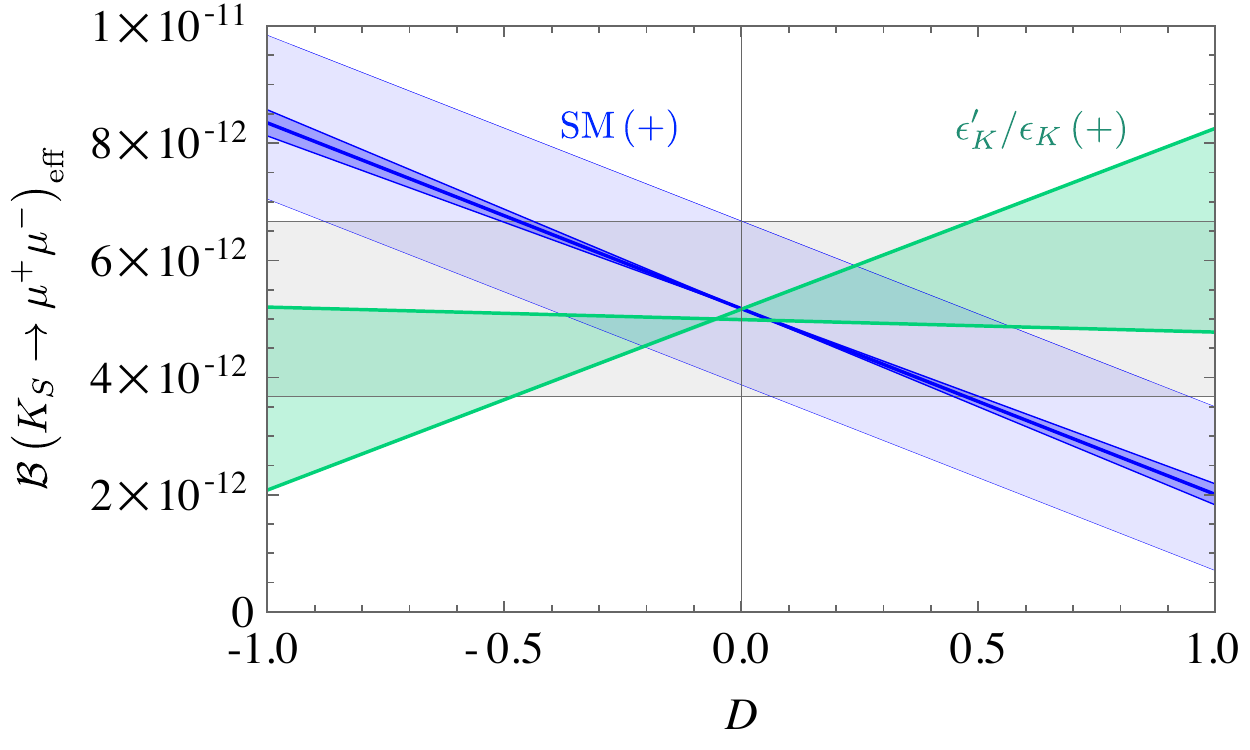}}
\end{minipage}
\caption[]{$K_S$   interference effect from eq. \ref{eq:effint} on $\mathcal{B} (K_S \to \mu^+ \mu^-)  $ depending on the  ${A}_{L\gamma\gamma}$ sign:  negative (center and in red SM while in green NP contributions)  and positive  (right and in blue SM while in green NP contributions).}
\label{fig:KSa}
\end{figure}

The LD contributions to  $K_{S}\to  \mu ^+ \mu ^-$ Fig. (\ref{fig:KSa})   have been  computed reliably in CHPT ($B=(5.0\pm 1.5)\times 10^{-12}$). The relevant short distance contributions are
  \beqa 
  B(K_S \rightarrow\mu\overline{\mu})_{SM}^{SD}&=1\times 10^{-5}& |\Im (V _{ts}^* V_{td})|^2\sim 10^{-13}  \nonumber\\
  B(K_S \rightarrow\mu\overline{\mu})_{NP}&\le 10^{-11}&\label{Hsd}\eeqa
  We have shown that in some appealing  susy scenario in Fig. (\ref{fig:KSsusy}) \cite{Chobanova:2017rkj}
large    allowed new physics contributions (NP)  can be substantially larger that SM SD contributions.

The short distance hamiltonian     will contribute also to  $K_L\rightarrow\mu\overline{\mu}$,  through a CP conserving amplitude,  $\Re (A_{\rm short})$, that 
has to be disentangled from the large LD two-photon exchange contributions, $A_{\gamma \gamma}$: the absorptive  LD contribution is   much larger than SD, in the rate respectively 25 times larger than dispersive; total $B_{\rm expt}=(6.84\pm 0.11)\times 10^{-9}$.
To extract SD info  the situation would be better if we would know the sign of  $A_{\gamma \gamma}$, theoretically and experimentally unknown. While $K_L -$decays outside the LHCB fiducial volume the interference $
A(K_L\rightarrow\mu\overline{\mu} ^* 
A(K_S\rightarrow\mu\overline{\mu}$ may affect the LHCB $K_S -$rates:
we can  study the time interference $K_{S,L}\to \mu \mu$ ; this can be done by flavor tagging  ${K^0}{\bar K}^0$  , specifically by detecting the associated
 $\pi ^\pm$ and (or)  $K^\mp$, determining the impurity parameter $D=\frac{K^0 -\bar{K}^0}{K^0 + \bar{K}^0}$. Then interference term will affect the measured branching \cite{DAmbrosio:2017klp}:
 \beqa
 \mathcal{B} (K_S^0 \to \mu^+ \mu^-)_{\rm eff}=\tau_S  \left( \int^{t_{{\rm max}}}_{t_{{\rm min}}} d t e^{- \Gamma_S t} \varepsilon (t)\right)^{-1}\nonumber\\
 \Biggl[ \int^{t_{{\rm max}}}_{t_{{\rm min}}} d t \Biggl\{  \Gamma (K^0_{S} \to \mu^+ \mu^- )  e^{- \Gamma_S t}  ~+   \frac{ D f_K^2 M_K^3 \beta_{\mu}}{ {8} \pi}\cdot\nonumber\\
 \textrm{Re}\left[  i \left( A_S A_L - \beta_{\mu}^2 {B_S^{\ast}} B_L \right) e^{ - i \Delta M_K t}  \right] e^{- \frac{ \Gamma_S + \Gamma_L}{2} t } \Biggr\} \varepsilon (t) \Biggr]\nonumber\\ \nonumber
\eeqa
 Then we are i) increasing the sensitivity to short distance and ii) possibly determining the sign ${A}_{L\gamma\gamma}$
\beqa
\sum_{\rm spin} \mathcal{A}(K_1 \to \mu^+ \mu^-)^{\ast} \mathcal{A}(K_2 \to \mu^+ \mu^- ) 
\sim \underbrace{\textrm{Im}[\lambda_t] y'_{7A}}_{SD}\nonumber\\ \left\{ \underbrace{{A}^{\mu}_{L\gamma\gamma}}_{LD} - 2 \pi \sin^2 \theta_W  \left(  \textrm{Re} [\lambda_t]  y'_{7A} + \textrm{Re} [\lambda_c]  y_c  \right) \right\}\label{eq:effint}
\eeqa
Experimentally, one can also access an {\it effective} branching ratio of $K_S^0 \to  \mu^+ \mu^-$~\cite{DAmbrosio:2017klp} which includes an interference contribution with $K_L^0 \to  \mu^+ \mu^-$ in the neutral kaon sample.
LHCB has a beautiful kaon physics program \cite{Chobanova:2017rkj,Junior:2018odx}.
 \section{The weak chiral lagrangian} 
In Ref. \cite{Cappiello:2017ilv} we have studied how to determine the weak O($p^4$) chiral countertems in 
\begin{equation}
\begin{array}{ll}
{ {\cal L}} _{\Delta S=1}={ {\cal L}} _{\Delta S=1} ^2 
+{ {\cal L}} _{\Delta S=1} ^4  +\cdots  \\ \\ \quad={{G_8} F^4 }
 \underbrace{
\la \lambda _6 
D_\mu U^\dag D^\mu U \rangle}_{{K \rightarrow 2 \pi}/3\pi } +
\underbrace{ {G_8 F^2 }\sum_i  N_i W_i}_{
K^+ \rightarrow \pi^+
 \gamma \gamma,{
{K \rightarrow \pi l^+ l^-} }}  \nonumber\end{array}\nonumber
\end{equation}
In fact as shown in Table I there is a subset of the 37 CT's,  $N_{14}^r, \  N_{15}^r, \  N_{16}^r$ and $N_{17}$, that can be determined from experiments.
Due to the accurate   NA48/2 study of the decays $K^{\pm}\to \pi^{\pm}\pi^0 \gamma$ and 
$K^{\pm}\to \pi^{\pm}\pi^0 e^+ e^-$ the subset of CT's in the table I can be determined
\begin{table*}[t]
\begin{center}
\caption{O($p^4$) weak chiral countertems and their determination}
\begin{tabular}{llcclc}
&$K^{\pm}\to \pi^{\pm}\gamma^*$& $N_{14}^r-N_{15}^r$&&$a_+=-0.578\pm 0.016$~&{\rm NA48/2}\\
&$K_S\to \pi^0\gamma^*$&$2N_{14}^r+N_{15}^r $&&$a_S=(1.06^{+0.26}_{-0.21}\pm 0.07)$~&{\rm NA48/1}\\
&$K^{\pm}\to \pi^{\pm}\pi^0\gamma$& $N_{14}^r-N_{15}^r-N_{16}^r-N_{17}$ &&$X_E=(-24\pm 4\pm 4)~
{\mathrm{GeV}}^{-4}$~&{\rm NA48/2}\\
&$K^+\to \pi^+\gamma\gamma$&$N_{14}^r-N_{15}^r-2N_{18}^r $&&${\hat{c}}=1.56\pm 0.23\pm 0.11$~&{\rm NA48/2}
\end{tabular}
 \end{center}
\end{table*}

\section{Lepton flavor universality violation in kaons}
The dominant contribution to $K^+\to\pi^+\ell^+\ell^-$ is due to single virtual-photon exchange. The   amplitude involves a vector form factor $V_+(z)$ which up to $O(p^6)$ in the chiral expansion, can be decomposed in the general form~\cite{DAmbrosio:1998gur} 
\begin{equation}
  V_+(z) = a_+ + b_+z + V_+^{\pi\pi}(z)\,,  \qquad z = q^2 / m_K^2\,.
  \label{V_dec}
\end{equation}
Here the LECs $a_+$ and $b_+$ parametrise the polynomial part, while the rescattering contribution $V_+^{\pi\pi}$ can be determined from fits to $K\to\pi\pi$ and $K\to\pi\pi\pi$ data.  Chiral symmetry alone does not constrain the values of the LECs,%
so instead, we consider the differential decay rate $d \Gamma / dz \propto |V_+(z)|^2$ as a means to extract $a_+$ and $b_+$ from experiment.  The resulting fit to the decay spectra from all available high-statistics experiments is given in Table~\ref{tab:a+b+}.

\begin{table}[t]
\renewcommand{\arraystretch}{1.3}
\centering
\begin{tabular}{cccr}\toprule 
Channel & $a_+$ & $b_+$ & Reference \\
\hline
$ee$ & $-0.587\pm 0.010$ & $-0.655\pm 0.044$ & E865~\cite{Appel:1999yq}\\
$ee$ & $-0.578\pm 0.016$ & $-0.779\pm 0.066$ & NA48/2~\cite{Batley:2009aa}\\
$\mu\mu$ & $-0.575\pm 0.039$ & $-0.813\pm 0.145$ & NA48/2~\cite{Batley:2011zz}\\
\hline
\end{tabular}
\caption{Fitted values of coefficients in the vector form factor.}
\label{tab:a+b+}
\end{table}

Now for the crucial point: if lepton flavour universality applies, the coefficients $a_+$ and $b_+$ have to be equal for the $ee$ and $\mu\mu$ channels, which within errors is indeed the case.  Since the SM interactions are lepton flavour universal, deviations from zero in differences like $a_+^{\mu\mu} - a_+^{ee}$ would then be a sign of NP, and the corresponding effect would be necessarily short-distance \cite{Crivellin:2016vjc}. 

To convert the allowed range on $a_+^{\rm NP}$ into a corresponding range in the Wilson coefficients $C_{7V}^{\ell\ell}$, we make use of the $O(p^2)$ chiral realization of the $SU(3)_L$ current
 $$
  \bar{s}\gamma^\mu(1-\gamma_5)d \leftrightarrow i F_\pi^2 (U\partial^\mu U^\dagger)_{23}\,, 
  \qquad U = U(\pi,K,\eta)\,,  
  $$
to obtain 
\begin{equation}
 a_+^{\rm NP}=\frac{2\pi\sqrt{2}}{\alpha}V_{ud}V_{us}^*C_{7V}^{\rm NP}\,.
 \label{aNP}
\end{equation}
Contributions due to NP in $K^+ \to \pi^+ \ell^+\ell^-$ can then be probed by considering the \emph{difference} between the two channels
\begin{equation}
\label{limit_Kp}
 C_{7V}^{\mu\mu}-C_{7V}^{ee}=\alpha\frac{a_+^{\mu\mu}-a_+^{ee}}{2\pi\sqrt{2}V_{ud}V_{us}^*}\,.
\end{equation}
If the framework of MFV, this can be converted into a constraint on the NP contribution to $C_9^B$:
\begin{equation}
\label{C_charged}
 C_{9}^{B,\mu\mu}-C_{9}^{B,ee}=-\frac{a_+^{\mu\mu}-a_+^{ee}}{\sqrt{2}V_{td}V_{ts}^*}\approx -19\pm 79\,,
\end{equation}
where we have averaged over the two electron experiments listed in Table~\ref{tab:a+b+}.

Evidently, the determination of $a_+^{\mu\mu}-a_+^{ee}$ needs to be improved by an $O(10)$ factor in order to probe the parameter space relevant for the $B$-anomalies, whose explanation involves Wilson coefficients $C_{9,10}^B=O(1)$~\cite{Descotes-Genon:2015uva}. Improvements of this size may be possible at NA62, especially for the experimentally cleaner dimuon mode which currently has the larger uncertainty.

\section{Determining $a_+ $ and  $b_+ $}
\begin{figure} 
\begin{center}
\includegraphics[width=3.5cm]{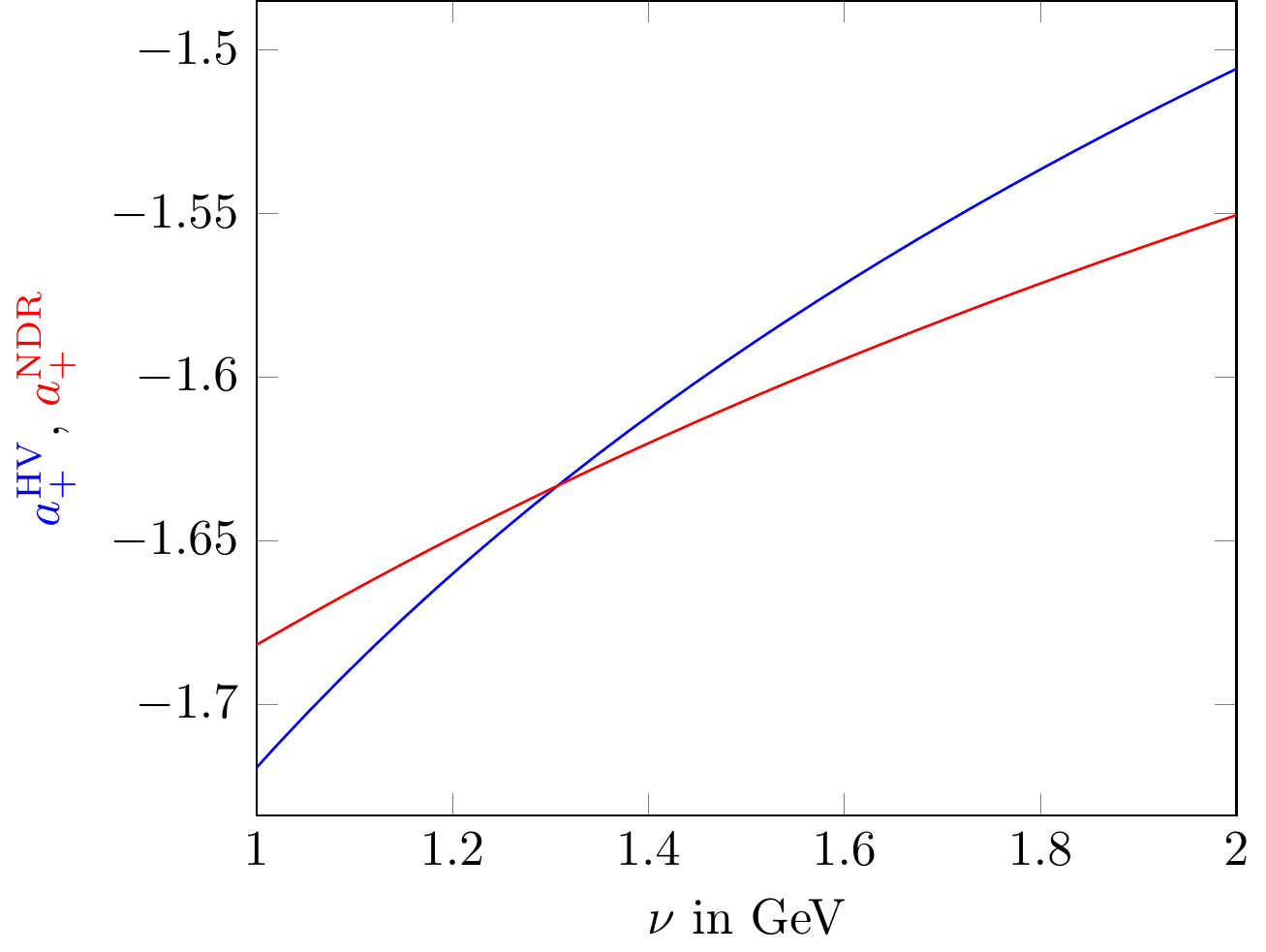} ~\hspace*{0.2cm}~ \includegraphics[width=3.5cm]{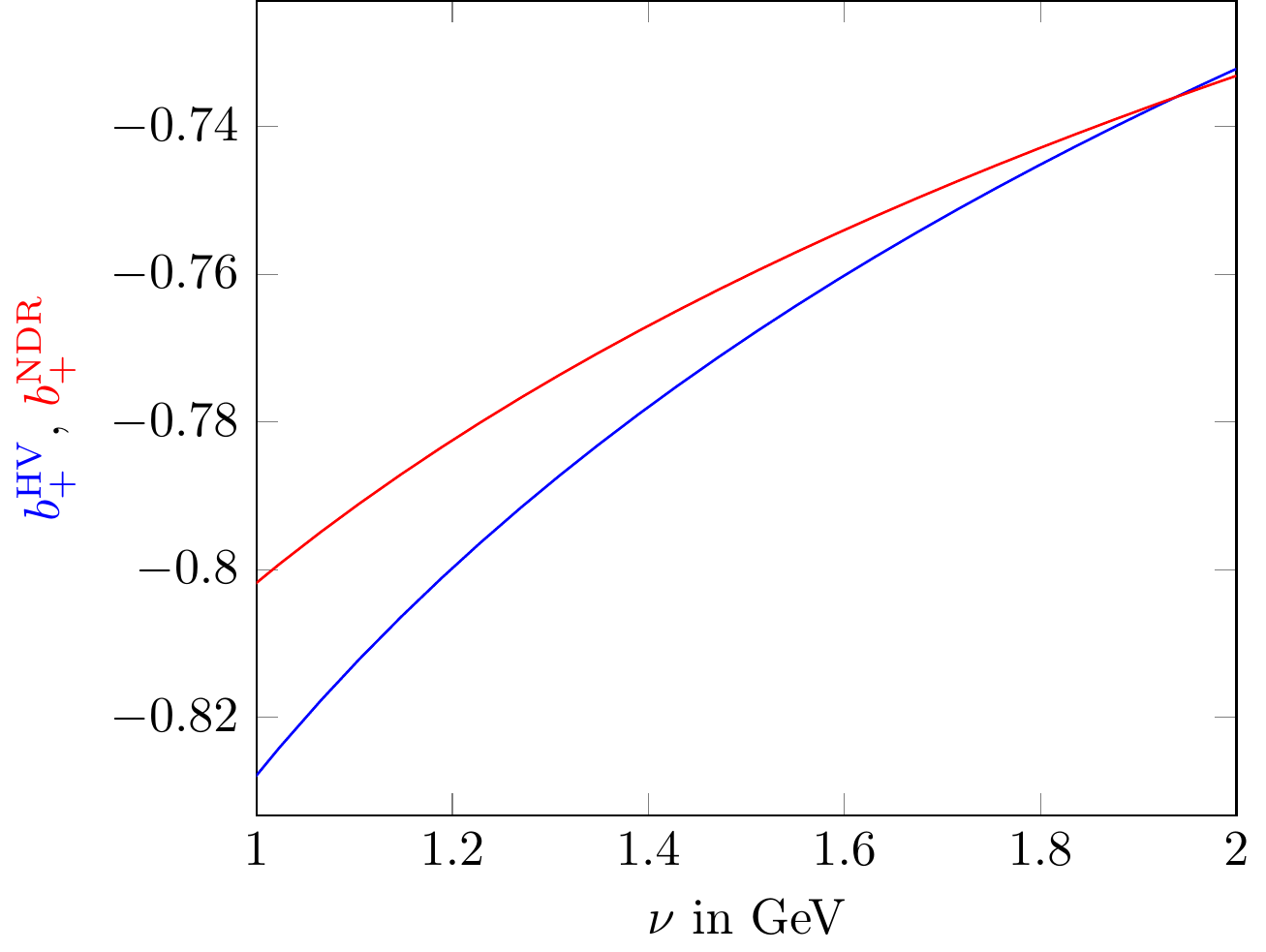}
\end{center}
\caption{  The evolution of  $a_+$ and $b_+$ with respect to $\nu$ in both NDR and HV schemes, for $M=1$ GeV. \label{Fig:11}}
\end{figure}

Recently we have addressed determining $a_+ $ and  $b_+ $ from low energy data  and short distance constraints\cite{DAmbrosio:2018ytt,DAmbrosio:2019xph}.
This is done by  first part describes the contribution from
the two-pion intermediate state to $V_+(z)$ in eq. \ref{V_dec}. It is constructed upon assuming, 
in analogy with the electromagnetic form factor of the pion $F_V^{\pi} (s)$
 that it is given by an unsubtracted dispersion integral,
\be
V_+^{\pi\pi} (z) = \int_{4 M_\pi^2}^\infty dx \frac{\rho_+^{\pi\pi} (x)}{x-zM_K^2-i0}
.
\label{W_pi-pi_disp}
\ee
The absorptive part consists of the two-pion spectral density $\rho_+^{\pi\pi} (s)$,
and is obtained upon inserting a two-pion intermediate state in the 
representation of the form factor given in eq. \ref{V_dec},
\bea
\rho_+^{\pi\pi} (s)& = 
16 \pi^2 M_K^2 \times \frac{s - 4 M_\pi^2}{s} \, \theta ( s - 4 M_\pi^2 )
\times
F_V^{\pi } (s) 
\times\nonumber\\
&\frac{  {f}^{K^\pm \pi^\mp \to \pi^+ \pi^-}_1 \!(s)}{\lambda^{1/2}_{K\pi} (s)}
.
\label{disc_FF_pi-pi_P-wave}
\eea
Data provide the scattering amplitude ${f}^{K^\pm \pi^\mp \to \pi^+ \pi^-}_1$ (fixing also the correct position of the $\rho$-poles !!), while matching short distance and resonances furnish also the remaining contributions  in eq. \ref{V_dec}.
Our predictions are in Fig.  \ref{Fig:11} \cite{DAmbrosio:2018ytt} and they have been improved in \cite{,DAmbrosio:2019xph}.
\begin{acknowledgments}
I would like to thank  the organizers of FCPC 2019 in particular Bob   Kowalewski.  I  also acknowledge collaboration with L. Cappiello, O. Cata, D. Greynat A. Iyer, T. Kitahara, M. Knecht, D.Martinez Santos and K. Yamamoto.
\end{acknowledgments}

\bigskip 

\begin{thebibliography}{9}   




\bibitem{Crivellin:2017gks} 
  A.~Crivellin, G.~D'Ambrosio, T.~Kitahara and U.~Nierste,
  Phys.\ Rev.\ D {\bf 96}, no. 1, 015023 (2017)
  doi:10.1103/PhysRevD.96.015023
  [arXiv:1703.05786 [hep-ph]].

 \bibitem{Mirra} Michal ZAMKOVSKY, these Proceedings  

\bibitem{Kitahara:2016otd} 
  T.~Kitahara, U.~Nierste and P.~Tremper,
  Phys.\ Rev.\ Lett.\  {\bf 117}, no. 9, 091802 (2016)
  doi:10.1103/PhysRevLett.117.091802
  [arXiv:1604.07400 [hep-ph]].

  

\bibitem{Hurth:2017sqw} 
  T.~Hurth, C.~Langenbruch and F.~Mahmoudi,
  JHEP {\bf 1711}, 176 (2017)
  doi:10.1007/JHEP11(2017)176
  [arXiv:1708.04474 [hep-ph]].



\bibitem{DAmbrosio:2017wis} 
  G.~D'Ambrosio and A.~M.~Iyer,
  Eur.\ Phys.\ J.\ C {\bf 78}, no. 6, 448 (2018)
  doi:10.1140/epjc/s10052-018-5915-9
  [arXiv:1712.08122 [hep-ph]].
 
\bibitem{Chakraborty:2017mbz}
  A.~Chakraborty, A.~M.~Iyer and T.~S.~Roy,
  arXiv:1707.07084 [hep-ph].

\bibitem{DAmbrosio:2017klp} 
  G.~D'Ambrosio and T.~Kitahara, Phys.\ Rev.\ Lett.\  {\bf 119}, no. 20, 201802 (2017)

\bibitem{Chobanova:2017rkj} 
  V.~Chobanova, G.~D'Ambrosio, T.~Kitahara, M.~Lucio Martinez, D.~Martinez Santos, I.~S.~Fernandez and K.~Yamamoto,
  JHEP {\bf 1805}, 024 (2018)
doi.org/10.1007/JHEP05(2018)024
  [arXiv:1708.04474 [hep-ph]]
  arXiv:1711.11030 [hep-ph].

\bibitem{Junior:2018odx} 
  A.~A.~Alves Junior {\it et al.},
  JHEP {\bf 1905}, 048 (2019)
  doi:10.1007/JHEP05(2019)048
  [arXiv:1808.03477 [hep-ex]].

\bibitem{Cappiello:2017ilv} 
  L.~Cappiello, O.~Cat\`a and G.~D'Ambrosio,
  Eur.\ Phys.\ J.\ C {\bf 78}, no. 3, 265 (2018)
  doi:10.1140/epjc/s10052-018-5748-6
  [arXiv:1712.10270 [hep-ph]].



\bibitem{DAmbrosio:1998gur} 
  G.~D'Ambrosio, G.~Ecker, G.~Isidori and J.~Portoles,
  JHEP {\bf 9808}, 004 (1998)
  doi:10.1088/1126-6708/1998/08/004
  [hep-ph/9808289].


\bibitem{Appel:1999yq}
  R.~Appel {\it et al.} [E865 Collaboration],
  Phys.\ Rev.\ Lett.\  {\bf 83} (1999) 4482
  [arxiv:hep-ex/9907045].


\bibitem{Batley:2009aa}
  J.~R.~Batley {\it et al.} [NA48/2 Collaboration],
  Phys.\ Lett.\ B {\bf 677} (2009) 246
  [arxiv:hep-ex/0903.3130].


\bibitem{Batley:2011zz}
  J.~R.~Batley {\it et al.} [NA48/2 Collaboration],
  Phys.\ Lett.\ B {\bf 697} (2011) 107
  [arxiv:hep-ex/1011.4817].


\bibitem{Crivellin:2016vjc}
  A.~Crivellin, G.~D'Ambrosio, M.~Hoferichter and L.~C.~Tunstall,
  Phys.\ Rev.\ D {\bf 93} (2016) 074038
  [arxiv:hep-ph/1601.00970].

\bibitem{Descotes-Genon:2015uva} 
  S.~Descotes-Genon, L.~Hofer, J.~Matias and J.~Virto,
  JHEP {\bf 1606}, 092 (2016)
  [arxiv:hep-ph/1510.04239].
 
\bibitem{DAmbrosio:2018ytt} 
  G.~D'Ambrosio, D.~Greynat and M.~Knecht,
  JHEP {\bf 1902}, 049 (2019)
  doi:10.1007/JHEP02(2019)049
  [arXiv:1812.00735 [hep-ph]].


\bibitem{DAmbrosio:2019xph} 
  G.~D'Ambrosio, D.~Greynat and M.~Knecht,
  arXiv:1906.03046 [hep-ph].
 

\end{thebibliography}

\end{document}